\documentclass[preprintnumbers,amsmath,amssymb,showkeys]{revtex4} 

\usepackage{graphicx}
\usepackage{dcolumn}
\usepackage{bm}

\begin{document}

\preprint{Pramana- j. of physics/DOI:10.1007/s12043-012-0323-4}

\title{Non-adiabatic radiative collapse of a relativistic star under different initial conditions}

\author{R. Sharma}
\email{rsharma@iucaa.ernet.in}
\affiliation{Department of Physics, P. D. Women's College, Jalpaiguri 735101, India.}

\author{R. Tikekar}
\email{tikekar@gmail.com}
\affiliation{Formerly at: Department of Mathematics, Sardar Patel University, Vallabh Vidyanagar, Gujarat, India.}
          
\date{\today}

\begin{abstract}
We examine the role of space-time geometry in the non-adiabatic collapse of a star dissipating energy in the form of radial heat flow, studying its evolution under different initial conditions. The collapse of a star with interior comprising of a homogeneous perfect fluid is compared with that of a star filled with inhomogeneous imperfect fluid with anisotropic pressure. Both the configurations are spherically symmetric, however, in the latter case, the physical space $t= constant$ of the configurations is assumed to be inhomogeneous endowed with spheroidal or pseudo-spheroidal geometry. It is observed that as long as the collapse is shear-free, its evolution depends only on the mass and size of the star at the onset of collapse.

\end{abstract}


\maketitle

\section{\label{sec:level1}Introduction}
Study of the evolution of a non-adiabatically collapsing spherical compact star under the influence of its self-gravity in the presence of dissipative forces and determining the final state it attains, is a field of active interest in theoretical astrophysics. The \citet{Vaidya} metric appropriately describes the exterior space-time background of such collapsing stars. It has been generally expected that various factors contributing to the space-time geometry and matter-content within it, such as shear, inhomogeneity, local anisotropy, presence of electromagnetic field etc., may directly affect the evolution of the collapse. This is not surprising since the Einstein's field equations $G_{\alpha\beta} = 8\pi T_{\alpha\beta}$ clearly imply a mutual correspondence between the composition of a gravitating body and geometry associated with its space-time.

\citet{Baner} had presented a metric describing space-time of a spherically symmetric, shear-free, collapsing object in the presence of heat flux. Sch$\ddot{a}$fer and Goenner\cite{Dirk} used this space-time metric and examined various features of the evolution of the collapse of a shear-free homogeneous and isotropic configuration of matter. \citet{Chan1,Chan2} analyzed the collapse of a matter distribution of uniform matter density with shear. We have examined the collapse of a relativistic star with matter density and fluid pressure decreasing radially outward from the centre, considering the background  space-time as having geometry different from that of a uniform density star. The space-time is spherically symmetric and it's associated $t= constant$ hyper surfaces have the geometry of either a $3$-spheroid or pseudo-spheroid. The use of space-time geometry of this type to obtain relativistic models of compact stars with no reliable information about their matter content was first proposed by \citet{Vaidya2}. Subsequently several physically viable models of realistic stars obtained adopting this approach have been discussed by many authors (e.g., see Ref.~\citep{Knutsen}-\citep{Sarwe}).  

In this paper, we have examined the impact of inhomogeneity of background space-time and anisotropic stresses on the collapse rate of a relativistic star which has lost its equilibrium. In Section \ref{sec:level2}, the field equations for a spherically symmetric anisotropic fluid undergoing non-adiabatic radiative collapse on the background of a spherically symmetric inhomogeneous space-time have been formulated and the implications of prescribing the $3$-spheroidal or $3$-pseudo-spheroidal geometry for the interior physical 3-space of the collapsing configuration have been examined. In Section \ref{sec:level3}, the nature of the space-time outside the collapsing star is discussed, the conditions to be fulfilled across the boundary surface separating it from the interior space-time have been formulated and an outline of the procedure for studying the evolution of the collapse is given.  The concluding Section \ref{sec:level4} is devoted to discussion of our observations about the impact of the inhomogeneity on the evolution of the collapse of compact stars.

\section{\label{sec:level2}Interior space-time}
We consider a compact spherical star in equilibrium with interior space-time described by metric
\begin{equation}
ds_{s}^2 = -A_0^2(r) dt^2 + B_0^2(r) dr^2 + r^2(d\theta^2 + \sin ^2 \theta d\phi ^2), \label{eq1}
\end{equation}
wherein the coordinates have been labeled as $x^0=t$, $x^1=r$, $x^2=\theta$, $x^3=\phi$. We assume that the interior content of the star is an imperfect fluid with energy-momentum tensor
\begin{equation}
T_{\alpha\beta} = [(\rho)_s + (p_t)_s] u_\alpha u_\beta +( p_t )_sg_{\alpha\beta} + [(p_r)_s - (p_t)_s]\chi_{\alpha} \chi_{\beta}. \label{eq2}
\end{equation}
In Eq.~(\ref{eq2}),  $(\rho)_s$ denotes the energy density, $(p_r)_s$ and $(p_t)_s$ respectively denote fluid pressures along the radial and transverse directions, $u^{\alpha}$ is the four velocity of the fluid, while $\chi^{\alpha}$ is a unit space-like four vector along the radial direction. The Einstein's field equations 
\begin{equation}
R_{\alpha\beta} -\frac{1}{2}g_{\alpha\beta}R = 8\pi T_{\alpha\beta}, \label{eq3}
\end{equation}
in system of units rendering $G = c = 1$, determine these dynamical parameters in terms of the metric coefficient functions as
\begin{eqnarray}
8\pi (\rho)_s &=&  \frac{1}{r^2}-\frac{1}{r^2 B_0^2}-\frac{2B'_0}{r B_0^3}, \label{eq4} \\
8\pi (p_r)_s &=&  -\frac{1}{r^2} + \frac{1}{B_0^2 r^2}+\frac{2 A'_0}{rA_0 B_0^2},\label{eq5}  \\  
8\pi (p_t)_s &=& 8\pi \left[(p_r)_s + {\Delta}_s(r)\right], \label{eq6} \\
8\pi {\Delta}_s &=& \frac{A''_0}{A_0 B_0^2}-\frac{A'_0}{rA_0 B_0^2}-\frac{B'_0}{rB_0^3}-\frac{A'_0 B'_0}{A_0 B_0^3} - \frac{1}{B_0^2 r^2} +\frac{1}{r^2}.\label{eq7}  
\end{eqnarray}
In above  ${\Delta}_s(r)$ represents the parameter which is a measure of anisotropy of the stresses. The subscript `s' is used to denote the dynamical variables associated with the compact star in equilibrium state. The interior space-time of the star after it begins to collapse on loosing its equilibrium will be described in general by the metric
\begin{equation}
ds_{-}^2 = -A^2(r,t) dt^2 + B^2(r,t) dr^2 + C^2(r,t)(d\theta^2 + \sin ^2 \theta d\phi ^2). \label{eq8}
\end{equation}
The interior matter of the star will be accompanied with radial flow of heat as it shrinks in size and the energy-momentum tensor associated with it will be  
\begin{equation}
T_{\alpha\beta} = (\rho + p_t) u_\alpha u_\beta + p_t g_{\alpha\beta} + (p_r - p_t)\chi_\alpha \chi_\beta +
q_\alpha u_\beta + q_\beta u_\alpha, \label{eq9}
\end{equation}
where $q^{\alpha}$ is the heat flux vector. Einstein field equations (\ref{eq3}) governing the evolution of the collapse is a highly non-linear system of second order partial differential equations. We adopt separability of metric variables which facilitates examining their implications. Without any loss of generality we couch the space-time metric of Eq.~(\ref{eq8}) in the form 
\begin{equation}
ds_{-}^2 = -A_0^2(r) dt^2 + R^2(t) B_0^2(r) dr^2 +  r^2 Y^2(t)(d\theta^2 + \sin ^2 \theta d\phi ^2), \label{eq10}
\end{equation}
and express the system of five field equations in the following form:
\begin{eqnarray}
8\pi \rho &=& \frac{1}{R^2} 8\pi (\rho)_s + \frac{1}{A_0^2}\left(\frac{\dot{Y}^2}{Y^2}+2\frac{\dot{R}\dot{Y}}{RY}\right) + \frac{1}{r^2}\left(\frac{1}{Y^2} - \frac{1}{R^2}\right), \label{eq11} \\
8\pi p_r &=& \frac{1}{ R^2}8\pi (p_r)_s - \frac{1}{A_0^2}\left(\frac{\dot{Y}^2}{Y^2}+2\frac{\ddot{Y}}{Y}\right) -\frac{1}{r^2}\left(\frac{1}{Y^2}-\frac{1}{R^2}\right),\label{eq12}  \\  
8\pi p_t &=& \frac{1}{R^2}8\pi (p_t)_s - \frac{1}{A_0 ^2}\left(\frac{\dot{R}\dot{Y}}{RY}+\frac{\ddot{R}}{R}+\frac{\ddot{Y}}{Y}\right),\label{eq13}  \\
8\pi \Delta (r,t) &=& \frac{1}{R^2} 8\pi {\Delta}_s +\frac{1}{A_0^2}\left(\frac{\dot{Y}^2}{Y^2} +\frac{\ddot{Y}}{Y}-\frac{\ddot{R}}{R} -\frac{\dot{R}\dot{Y}}{RY}\right)+ \frac{1}{r^2}\left(\frac{1}{Y^2} - \frac{1}{R^2}\right),\label{eq14} \\
8\pi q &=& - \frac{2}{A_0 B_0^2}\left(\frac{A'_0\dot{Y}}{A_0 R^2 Y}+\frac{\dot{R}}{r R^3}-\frac{\dot{Y}}{r R^2 Y}\right). \label{eq15}
\end{eqnarray}
In Eq.~(\ref{eq14}), the function $\Delta(r,t)$ represents evolution of the anisotropy in the collapsing configuration. The form of this function indicates that the evolution of anisotropy is linked with presence of shear  ($Y \neq R$) in the background. Even a compact star with a perfect fluid interior with vanishing ${\Delta}_s(r)$ which starts shrinking in size when its equilibrium is lost will subsequently witness evolution of anisotropy in stresses as collapse proceeds. Apparently, shear is responsible for the development of anisotropic stresses in collapsing matter. 

The metric functions in Eq.~(\ref{eq1}) describing the interior space-time of a compact star in equilibrium are determined by solving the appropriate set of Einstein's field equations which needs reliable information about the equation of state of the matter content to be furnished a priori. For superdense stars with matter density exceeding nuclear density, this information is not available with certainty. In the absence of such information we adopt the alternative approach  suggested by \citet{Vaidya2} of prescribing a suitable geometry for the physical 3-space of the star and write the metric describing the interior space-time of the collapsing star as
\begin{equation}
ds_{-}^2 = - A_0(r)^2dt^2 + R(t)^2\left(\frac{1+\lambda k r^2}{1 - k r^2}\right)dr^2 + r^2R^2(t)(d\theta^2 + \sin ^2 \theta d\phi ^2), \label{eq16}
\end{equation}
where, we have set $R(t) =Y(t)$. The geometry of the physical space $t=constant$ hyper surface of the space-time described by (\ref{eq16}) is (i) spheroidal for $k= +$ve, and (ii) pseudo-spheroidal for $k= -$ve. \citet{Mukherjee} and \citet{Tikekar3} have shown that the respective matter distributions are physically plausible if $\lambda > 3/17$ and $\lambda < -1$. The constant $\lambda$ is the measure of departure from homogeneous 3-spherical geometry of a uniform density star. When $k = 0$, the physical 3-space is a flat Euclidean space. 
Substitution  
\begin{equation}
B_0 = \sqrt{\frac{1+\lambda k r^2}{1-k r^2}},\label{eq17}\\
\end{equation}
in Eq.~(\ref{eq7}) puts it in the form of a second order differential equation 
\begin{equation}
(1+\lambda k-\lambda k x^2)\frac{d^2 A_0}{dx^2} + \lambda k x \frac{d A_0}{dx} + \left[\lambda k(\lambda k+1)-\frac{(1+\lambda k-\lambda  k x^2)^2{8\pi \Delta}_s}{k(1-x^2)}\right]A_0= 0, \label{eq18}
\end{equation}
determining the metric parameter $A_0(r)$ where, $x$ denotes new independent variable defined as $x^2=1-k r^2$. In terms of a new dependent variable
\begin{equation}
\Psi(x) = A_0(x)(1+\lambda k-\lambda k x^2)^{-1/4},\label{eq19}
\end{equation}
Eq.~(\ref{eq18}) assumes the form
\begin{equation}
\frac{d^2\Psi}{d x^2} + \left[\frac{2\lambda k(\lambda k+1)(2\lambda k+1)-(4\lambda k+7)\lambda^2 k^2 x^2}{4(1+\lambda k-\lambda  k x^2)^2}-\frac{(1+\lambda k-\lambda k x^2){8\pi \Delta}_s}{k(1-x^2)}\right]\Psi = 0.\label{eq20}
\end{equation}
$\Psi(x)$ can be determined from Eq.~(\ref{eq20}) only when the nature of anisotropic parameter $\Delta_s$ is known in the interior of the star. It is expected that $\Delta_s$ should be regular everywhere in the interior. The equation (\ref{eq20}) admits a simple tractable regular solution
\begin{equation}
\Psi = C + D x,\label{eq21}
\end{equation}
with $C$ and $D$ as arbitrary constants for 
\begin{equation}
{\Delta}_s = \frac{k(1-x^2)[2\lambda k(\lambda k+1)(2\lambda k+1)-(4\lambda k+7)\lambda^2 k^2 x^2]}{32\pi(1+\lambda k-\lambda k x^2)^3}.\label{eq22}
\end{equation}
The above expression for ${\Delta}_s$ is regular at all interior points. The metric function $A_0(r)$ accordingly is explicitly found to be 
\begin{equation}
A_0(r) = (1+\lambda k^2 r^2)^{1/4} (C +D\sqrt{1-k r^2}).\label{eq23}
\end{equation}
Subsequently the interior metric for space-time of the shear-free collapsing star with matter will be
\begin{equation}
ds_{-}^2 = -(1+\lambda k^2 r^2)^{1/2}(C +D\sqrt{1-k r^2})^2 dt^2 + R^2(t)\left[\frac{1+\lambda k r^2}{1-k r^2}\right] dr^2 + r^2R^2(t)(d\theta^2 + \sin ^2 \theta d\phi ^2). \label{eq24}
\end{equation}
Space-time of Eq.~(\ref{eq24}) is an inhomogeneous generalization of the  space-time of the metric of Banerjee {\em et al}\cite{Baner} 
\begin{equation} 
ds_{-}^2 = -(C -\sqrt{1-k r^2})^2 dt^2 + R^2(t)\left[\frac{dr^2}{1-k r^2} + r^2(d\theta^2 + \sin ^2 \theta d\phi ^2)\right], \label{eq25}
\end{equation}
as it can be recovered on setting $\lambda = 0$ and $D=-1$ from it.  

The progress of the collapse is directly linked with the metric function $R(t)$ which plays the role of scale factor as the configuration shrinks in size. The determination of $R(t)$ is discussed in the following Section.

\section{\label{sec:level3}Exterior space-time and junction conditions}
Since the non-adiabatic collapse of the stellar configuration is accompanied with heat flux the space-time in its exterior is expected to witness presence of radiation and will be appropriately described by the Vaidya metric
\begin{equation}
ds_{+}^2 = -\left(1-\frac{2m(v)}{\bar{r}}\right)dv^2 - 2dvd\bar{r} + \bar{r}^2[d \theta^2 + \sin^2 \theta d \phi^2],\label{eq26} 
\end{equation}
with $m(v)$ and $\Sigma$ denoting respectively the total mass and the shrinking boundary of the star, both functions of the retarded time $v$. 

The metrics of the interior and the exterior space-time of the star should join across its boundary surface - a $2$-sphere of the evolving system. This is done by using the set of boundary conditions for a spherical collapsing matter distribution in the presence of heat flux completely formulated by Santos\cite{Santos} which ensure continuity of the first and second fundamental forms of the space-times. These conditions imply the following relations: 
\begin{eqnarray}
A_0(r_{\Sigma})dt &=& \left(1-\frac{2m}{\bar{r}}+2\frac{d{\bar{r}}}{dv}\right)^{\frac{1}{2}}_{\Sigma} dv, \label{eq27}\\
r_{\Sigma}R(t) &=& \bar{r}_{\Sigma}(v), \label{eq28}\\
m(v) &=& \frac{r_{\Sigma}R(t)}{2}\left[1 -\frac{1}{B_0^2}+\left(\frac{r\dot{R}}{A_0}\right)^2\right]_{\Sigma},\label{eq29}\\
\left[p_r\right]_{\Sigma} &=& \left[q R(t)B_0\right]_{\Sigma},\label{eq30}\\
m(r,t) &\stackrel{\Sigma}{=}& m(v), \label{eq31}
\end{eqnarray}
where, the functions $B_0(r)$ and $A_0(r)$ are as given in Eqs.~(\ref{eq17}) and (\ref{eq23}), respectively. The mass function $m(r,t)$ defined as
\begin{equation}
m(r,t) = \frac{rR(t)}{2}\left[1 -\frac{1}{B_0^2}+\left(\frac{r\dot{R}}{A_0}\right)^2\right],\label{eq32}
\end{equation}
represents the total mass enclosed within the boundary surface $r\leq r_\Sigma$ at time $t$.
It is expected that before the collapse sets in there will be no heat flow across the boundary surface of the star. The corresponding boundary condition  implies that $(p_r)_s (r=r_{\Sigma}) = 0$.  The condition (\ref{eq30}) leads to a differential equation
\begin{equation}
2\ddot{R}R + \dot{R}^2 - 2n\dot{R} = 0, \label{eq33}
\end{equation}
where, 
\begin{equation}
n = \left[\frac{A'_0}{B_0}\right]_{\Sigma},\label{eq34}
\end{equation}
is a constant.
Following Bonnor {\em et al}\cite{Bonnor}, we write Eq.~(\ref{eq33}) as a first order differential equation
\begin{equation}
\dot{R} = -\frac{2n}{\sqrt{R}}(1-\sqrt{R}), \label{eq35}
\end{equation}
which admits 
\begin{equation}
t=\frac{1}{n}\left[\frac{R}{2}+\sqrt{R} +ln(1-\sqrt{R})\right],\label{eq36}
\end{equation} 
as its solution. It follows that at $t \rightarrow -\infty$, the time when collapse is assumed to begin, $R = 1$ and $R=0$ at $t = 0$. Hence $R(t)$ will be  a decreasing function of time. 

Let $r_s$ be the radius of the star when it was in equilibrium and $m_s$  be the corresponding mass of the star so that $2m_s/r_s < 1$. When instability sets in, the super-massive star loses its equilibrium and begins to collapse. We denote this initial epoch when collapse sets in as  $t=-\infty$ so that $R(t)=1$. The comoving boundary surface $(rR)_{\Sigma} = r_s R(t)$ then starts shrinking until it reaches its Schwarzschild horizon value $\left[rR(t_{BH})\right]_{\Sigma} = r_s R(t_{BH}) = 2m(v)$, with $t=t_{BH}$ denoting the time of formation of the black hole. From Eq.~(\ref{eq29}), the mass of the evolving star at any instant will be
\begin{equation}
m(v) = \left[m_s R + \frac{2n^2r^3}{A_0 ^2}(1-\sqrt{R})^2\right]_{\Sigma},\label{eq37}
\end{equation} 
where, 
\begin{equation}
m_s = \frac{(\lambda+1)k r_s^3}{2(1+\lambda k r^2_s)}. \label{eq38} 
\end{equation} 
Consequently, the condition $(r R_{BH})_{\Sigma} = 2m(v)$ yields
\begin{equation}
R_{BH} = \left[\frac{\frac{2A_0' r}{A_0 B_0}}{\frac{2A_0' r}{A_0 B_0} +\sqrt{1-\frac{2m_s}{r}}}\right]^2_{\Sigma},\label{eq39}
\end{equation}
and the time of black hole formation is obtained as 
\begin{equation}
t_{BH} = \frac{1}{n}\left[\frac{R_{BH}}{2}+\sqrt{R_{BH}} +ln(1-\sqrt{R_{BH}})\right].\label{eq40}
\end{equation} 
We note that $R_{BH}$ depends on the initial values of the functions $A_0$ and $B_0$ on the boundary of the star when it starts moving inward as the collapse begins. The matter content of the star when the collapse begins is considered to be an inhomogeneous distribution of anisotropic matter which implies that $\lambda \neq 0$.  If $\lambda = 0,~ 0 < k < 1$, the star comprises of an incompressible distribution of perfect fluid with uniform matter density. The constants $C$, $D$, $k$ and $\lambda$ appearing in the functions $A_0(r)$ and $B_0(r)$ are determined by the boundary conditions which join the interior space-time metric of the star with the Schwarzschild exterior metric describing its exterior space-time, at the boundary of separation $r_s$. We stipulate the relevant boundary conditions as requiring continuity of metric tensor on $r_s$ as  
\begin{equation}
A_0(r_s) = \left(1-\frac{2m_s}{r_s}\right)^{1/2},~~~B_0(r_s) = \left(1-\frac{2m_s}{r_s}\right)^{-1/2}, \label{eq41}
\end{equation}
and vanishing of fluid pressure along radial direction $(p_r)_s(r_s) = 0$. These conditions determine
\begin{equation}
C = - \frac{D\sqrt{1-k r_s^2}(3+\lambda -\lambda k +4\lambda k^2 r_s^2 +\lambda^2 k^2 r_s^2)}{(1+\lambda -\lambda k +2\lambda k^2 r_s^2+\lambda^2 k^2 r_s^2)}. \label{eq42}
\end{equation} 
Note that $R_{BH}$ depends on the values of  $A_0(r)$ and $B_0(r)$ on the star boundary when collapse begins. The values of $A_0(r)$ and $B_0(r)$ on the star boundary are completely determined by the mass and size of the star at the onset of collapse. Hence $R_{BH}$ depends on the initial  mass and size of the star. Therefore, in this framework, the time of formation of black holes $t_{BH}$ and the radius $r_{BH} =r_s R_{BH}$ of the black hole remain the same for all stars of identical initial mass and size irrespective of the choice of $\lambda$.  We considered several stellar configurations in equilibrium of identical initial masses and radii (we assumed $m_s = 3.24~M_{\odot}$ and radius $r_s = 18~$km) with different values for the parameter $\lambda$ implying different compositions of interior stellar matter and obtained estimates of $R_{BH}$ using numerical procedures. We observed, as indicated in the Table \ref{tab:table1}, that all configurations yielded almost the same $R_{BH}$.

\begin{table*}
\caption{\label{tab:table1} Estimates of $R_{BH}$ for initial stellar configurations of mass $m_s = 3.24~M_{\odot}$ and radius $r_s = 18~$km with different compositions of interior stellar matter.}
\begin{ruledtabular}
\begin{tabular}{ccccc}
$\lambda$ & $C$ & $D$ & $k$ & $R_{BH}$ \\ 
5 & 2.73837 & -2.23899 & 0.00049 & 0.28196 \\
0 & 1.02725 & -0.49999 & 0.00164 & 0.28196 \\
-5 & -0.68817 & -1.16291 & 0.00122 & 0.28196 \\  
\end{tabular}
\end{ruledtabular}
\end{table*}

Accordingly, the parameter $\lambda$ seems to have no effect on the collapse rate. The inhomogeneous character of the interior matter content as implied by the geometry (spheroidal, pseudo-spheroidal or spherical) of interior physical space and nature of matter composition (fluid with isotropic or anisotropic stress structures) have no impact on the collapse rate of a stellar configuration. The function $R(t)$ is only a scaling parameter and carries no information about the geometry or material composition of the star in this set up. This is consistent with implications of Birkhoff's theorem according to which no observer located in the exterior region of a spherical star can get any information about the interior state of the star except its mass and size.

\section{\label{sec:level4}Discussion}
Considering the non-adiabatic collapse of a spherical star of uniform density Oliveira {\em et al}\cite{Oliveira1} have shown that some of the features like time required to form the black hole depend on the initial mass and size of the star. In our work, we have generalized the space-time metric of \citet{Baner} to investigate the effects of inhomogeneity and anisotropy on the collapse. Our study indicates that collapse rate for stars having identical initial mass and size does not depend on the geometry of their interior physical space. They take same time to collapse into their horizon and become black holes. Since the geometry of the physical space of a star in equilibrium is related with the composition of the interior matter of the star, this implies that stellar configurations having identical initial masses and radii irrespective of their composition, after loss of equilibrium, will take same time to become black hole.  

The analysis above is based on the following two assumptions. The space-time in the interior of the star (1) is spherically symmetric and shear-free, and (2) described by metric with metric functions as separable functions of space and time variables. Analysis based on such simplifying assumptions may give the same result unless new techniques are invoked to extract information about the change in the equation of state of the interior matter of a collapsing relativistic star and the mechanism for energy transport.  The mass and radius of a spherical star in equilibrium will be the only governing parameters which determine its collapse rate and time for its becoming a black hole after it begins to collapse unless shear develops in its space-time structure. Therefore, it is essential to consider non-adiabatic collapse of stellar configurations on the background of spherically symmetric space-time admitting shear $R(t) \neq Y(t)$ and study the impact of initial anisotropy and inhomogeneity on the evolution of the collapse.

\begin{acknowledgments}
RS and RT gratefully acknowledge support from IUCAA, Pune, India, where a part of this work was done. RT thanks the UGC of India for the award of Emeritus Fellowship.
\end{acknowledgments}

\end{document}